\documentclass{SpaceOps}
\PassOptionsToPackage{dvipsnames}{xcolor}
\RequirePackage[l2tabu, orthodox]{nag}

\usepackage{graphicx}
\usepackage{subcaption}
\usepackage{titlesec}
\usepackage{rotating}
\usepackage{array}

\renewcommand{\thesection}{\arabic{section}.} 
\renewcommand{\thesubsection}{\arabic{section}.\arabic{subsection}} 
\renewcommand{\thesubsubsection}{\arabic{section}.\arabic{subsection}.\arabic{subsubsection}} 

\titleformat{\section}[block]{\normalsize\bfseries}{\thesection}{0.5em}{}
\titleformat{\subsection}[block]{\normalsize\itshape}{\thesubsection}{0.3em}{}
\titleformat{\subsubsection}[block]{\normalsize\itshape}{\thesubsubsection}{0.3em}{}

\SpaceOpspaperyear{2025} 
\SpaceOpspapernumber{151} 
\SpaceOpslocation{Montreal, Canada} 
\SpaceOpsdate{26--30 May 2025} 

\SpaceOpscopyrightB{}

\title{Custody Transfer and Compressed Status Reporting for Bundle Protocol Version 7}

\SpaceOpsauthors{%
    Alice Le Bihan\textsuperscript{a*},
    Felix Flentge\textsuperscript{b},
    Juan A. Fraire\textsuperscript{cd}
}

\SpaceOpsaffiliations{%
    \textsuperscript{a} \emph{INSA Lyon, CITI, UR3720, 69621 Villeurbanne, France, \normalfont{\authormail{alice.le-bihan@insa-lyon.fr}}}\\
    \textsuperscript{b} \emph{Ground Systems Engineering and Innovation Department, ESA/ESOC, Darmstadt, Germany}\\
    \textsuperscript{c} \emph{Inria, INSA Lyon, CITI, UR3720, 69621 Villeurbanne, France}\\
    \textsuperscript{d} \emph{Universidad Nacional de Córdoba, Córdoba, Argentina}
}

\begin{document}

\abstract{
\indent As space missions increase, there is a growing need to replace point-to-point communication with an efficient and reliable network-centric communication approach. Disruption/Delay Tolerant Networking (DTN) with the Bundle Protocol (BP) has been selected as an interoperable network protocol in the LunaNet Interoperability Specification. It is also considered for future Earth Observation and Mars communication scenarios. In a DTN, the ‘bundle’ —the fundamental data unit of the Bundle Protocol— requires dedicated mechanisms to ensure reliability due to the inherent challenges posed by intermittent connectivity and long delays. The previous version of the Bundle Protocol, BPv6, contained a mechanism for reliable transfer between ‘custodial nodes’ called ‘custody transfer'. This mechanism moved responsibility for the reliable bundle delivery from one custodian node to the next and applied a time-based retransmission mechanism in case lost bundles were detected. However, this approach has been removed from the core protocol specification for the novel BPv7, which requires a corresponding Bundle Protocol reliability extension to be defined separately. This paper introduces a new custody transfer process for BPv7, which is expected to be published by CCSDS as an experimental specification in 2025. This new protocol extension harvests the ideas of Aggregate Custody Signalling but generalizes the concept to cover a broader range of possible usage scenarios. It is also based on concepts for efficient reporting on the status of sets of bundles, as custody acceptance or rejection can be seen as a case of more general status reporting. Specifically, the core features of this new custody transfer method for BPv7 are: (1) A strategy to efficiently identify sets of bundles by sequence numbering (2) A new Custody Transfer Extension Block and a corresponding administrative record, Compressed Custody Signal, to efficiently report on the acceptance or rejection of custody using the aforementioned sequence numbering (3) A new Compressed Reporting Extension Block requesting reporting on bundle reception, deletion, forwarding, delivery, or other processing steps using a corresponding administrative record with sequence numbering for efficiency. The paper will describe those concepts and their design, specification, and implementation in detail. These mechanisms have been prototyped in the ESA BP implementation and tested in Earth Observation and Lunar communication simulation scenarios. The results will be presented, as will an outlook on future work in the DTN reliable transfer domain.\\[6pt plus 2pt]

\newline

    \noindent
    \textbf{{Acronyms/Abbreviations:}} 
    
    ADU - Application Data Unit
    
    BP - Bundle Protocol
    
	CBOR - Concise Binary Object Representation  
            
	CRS - Compressed Reporting Signal
    
    CCS - Compressed Custody Signal
    
    CREB - Compressed Reporting Extension Block
    
	CTEB - Custody Transfer Extension Block

    E2E - End-to-end
    
	MIB - Management Information Base
            
	RTT - Round Trip Time

    NSE$^2$ - Network Simulation \& Emulation Environment    
}

\maketitle

\section{Introduction}
As space missions increase, there is a growing need to replace point-to-point communication with an efficient and reliable network-centric communication approach. Disruption/Delay Tolerant Networking (DTN) with the Bundle Protocol (BP) has been selected as an interoperable network protocol in the LunaNet Interoperability Specification~\cite{LunaNet}. It is also considered for future Earth Observation and Mars communication scenarios~\cite{FlentgeSpaceOps2021}. In a DTN, the ‘bundle’ —the fundamental data unit of the Bundle Protocol— requires dedicated mechanisms to ensure efficient reliability and retransmission due to the inherent challenges posed by intermittent connectivity and long delays. Reliability on individual links can be guaranteed by reliable link layer protocols such as the Licklider Transmission Protocol~\cite{CCSDSLTP}. However, it may not always be possible to apply those protocols as they require bi-directional communication. It is not always sufficient, as data may also be lost while it is at rest at one of the DTN nodes. End-to-end reliability can be achieved by application layer mechanisms. Still, those are not very efficient in typical relayed communication scenarios, e.g., if a bundle sent by a Mars rover via a Mars relay is lost on the trunk link back to Earth. In this case, retransmission must be triggered on the rover, which may only be possible much later due to intermittent connectivity. 

The previous version of the Bundle Protocol, BPv6~\cite{RfC5050}, contained a mechanism for reliable transfer between ‘custodial nodes’ called ‘custody transfer'. This mechanism moved responsibility for the reliable bundle delivery from one custodian node to the next and applied a timer-based retransmission mechanism in case lost bundles were detected. However, this approach has been removed from the core protocol specification for the novel BPv7~\cite{RFC9171} as estimating suitable timer values can be difficult in certain scenarios. Nevertheless, the custody transfer mechanism has been used widely in BPv6 deployments and is still deemed a required feature for BP-based DTNs. For this reason, the Bundle-in-Bundle Encapsulation IETF draft~\cite{BIBE} contained a retransmission mechanism initially called 'BIBE-custody'\footnote{In the meantime, this mechanism has been renamed to Bundle Retransmission mechanism (BRM).}.
In contrast to the original BPv6 Custody Transfer, this mechanism did require specifying the next custodian explicitly as the BIBE destination endpoint. Therefore, a new custody transfer based on the use of Bundle Protocol extension blocks and administrative records has been defined and will be described below. The 'new' BPv7 custody transfer provides more flexibility regarding retransmission mechanisms, so that others apart from timer-based retransmission can be used depending on the specific scenarios. Custody reporting has been inspired by the Aggregate Custody Signalling defined in the CCSDS Bundle Protocol standard~\cite{CCSDSBP}. 

The principles of Aggregate Custody Signalling can actually be extended and generalized to allow also efficient reporting of bundle status, such as delivery or deletion of bundles. Bundle Status Reporting has been defined in RFC 9171 but requires creating a single status report per bundle. This can create a huge amount of traffic, and for that reason, status reporting must be disabled per default and can only be used in cases where the related risk is deemed acceptable.
The Compressed Status Signals defined below can drastically reduce the amount of bundle status reporting-related traffic and can be used in operational scenarios. In particular, combining custody transfer with compressed delivery and deletion reporting is applicable to many space mission scenarios with limited communication opportunities but requirements for reliable data return.

In the following, we will first describe a way to uniquely identify 'sequences of bundles', essential for efficient reporting and custody signalling. Then, the BP extension blocks and the related BP administrative records for Custody Transfer and Compressed Bundle Reporting will be introduced. Initial experiments performed in simulated Lunar and Earth Observations scenarios demonstrate the effectiveness and efficiency of the proposed extensions. We conclude with a brief summary and the next steps.

\section{Custody Transfer and Compressed Bundle Reporting Extensions}
This section describes two protocol extensions for BPv7, which have been implemented and tested in ESA's  Bundle Protocol implementation. The first one, Compressed Bundle Reporting, aims to provide the same and additional reporting functionalities as Bundle Status Reporting as defined in RFC 9171~\cite{RFC9171}, but allowing to report on several bundles using only one Administrative Record \footnote{In BP, protocol-related information are carried by standard Application Data Units (ADUs) called Administrative Records.}, called Compressed Reporting Signal (CRS). This mechanism was previously introduced in~\cite{AidaManoleCBR}, and the model presented here is a continuation and improvement of this work. 
The second extension, Compressed Custody Signalling, is designed to provide all of the functionalities of Custody Transfer while summarising the custody status of several bundles in only one Compressed Custody Signal (CCS), therefore reducing the overall number of reports in the network.

\subsection{Uniquely identifying a bundle in the network} \label{unique_id}

Because Compressed Bundle Reporting and Compressed Custody Signalling both involve reporting on bundles, they share one fundamental pre-requirement: the ability to uniquely identify a bundle in the network. Unique bundle identification is possible on the bundle's source node ID and the bundle timestamp. However, for more efficient identification of multiple bundles, an additional mechanism is defined based on sequence numbering of bundles, which will be carried by the bundle itself using an extension block\footnote{In BP, a bundle is composed of a succession of blocks. The first one, the Primary Block, is mandatory. It carries fundamental information about the bundle, such as the version of BP used by its source node, its destination endpoint ID, its lifetime, etc. Then follows a list of extension blocks (one of which must be the Payload Block) that can be used to provide additional services in BP.}. For that purpose, two new extension blocks were introduced: the Compressed Reporting Extension Block (CREB) and the Custody Transfer Extension Block (CTEB), detailed respectively in Section~\ref{cbr_section} and Section~\ref{ccs_section}. The identification properties are assigned by the Bundle Node that inserts the extension block.

\begin{figure}[h]
    \centering
    \includegraphics[width=0.7\linewidth]{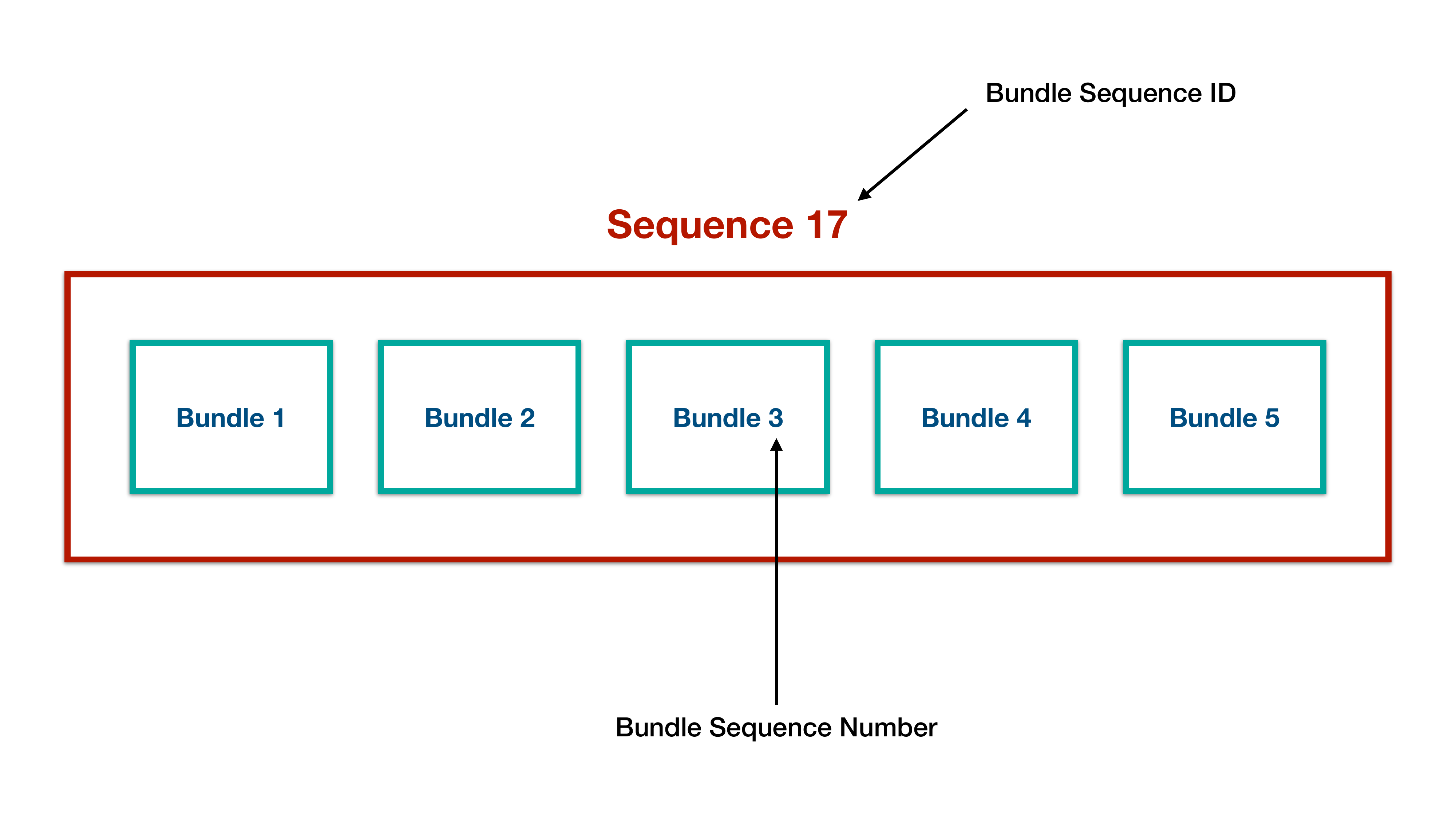}
    \caption{Example of a sequence of bundles.}
    \label{fig:sequence_id}
\end{figure}

The identification mechanism is based on sequences: each bundle with Compressed Custody Signalling or Compressed Bundle Reporting requested provides a sequence number qualified by a Bundle Sequence ID (see Figure~\ref{fig:sequence_id}). Using its Bundle Sequence ID and Bundle Sequence Number, the bundle can be uniquely identified within the node's scope. An additional element is needed to identify it throughout the whole network uniquely: the Administrative Endpoint ID of the node that inserted the extension block. A formal definition of each property is given below:

\begin{itemize}
    \item \textbf{Bundle Sequence ID}: each bundle with Compressed Bundle Reporting or Compressed Custody Signalling requested belongs to a sequence with a specific integer ID. The special ID ”0” is reserved, and its use will be explained later in this section.
    
    \item \textbf{Bundle Sequence Number}: each bundle in a sequence gets assigned a number between 0 and N, where N is a maximum value chosen so that no two bundles in a sequence receive the same Bundle Sequence Number during their lifetimes. Bundle Sequence Numbers are generated by objects called Bundle Sequence Counters, which increase the value by one every time a Bundle Sequence Number is used in an extension block.
    
    \item \textbf{Block Source Administrative Endpoint ID}: the Administrative Endpoint ID of the node that inserted the CTEB/CREB.
\end{itemize}

\begin{figure}
    \centering
    \includegraphics[width=0.9\linewidth]{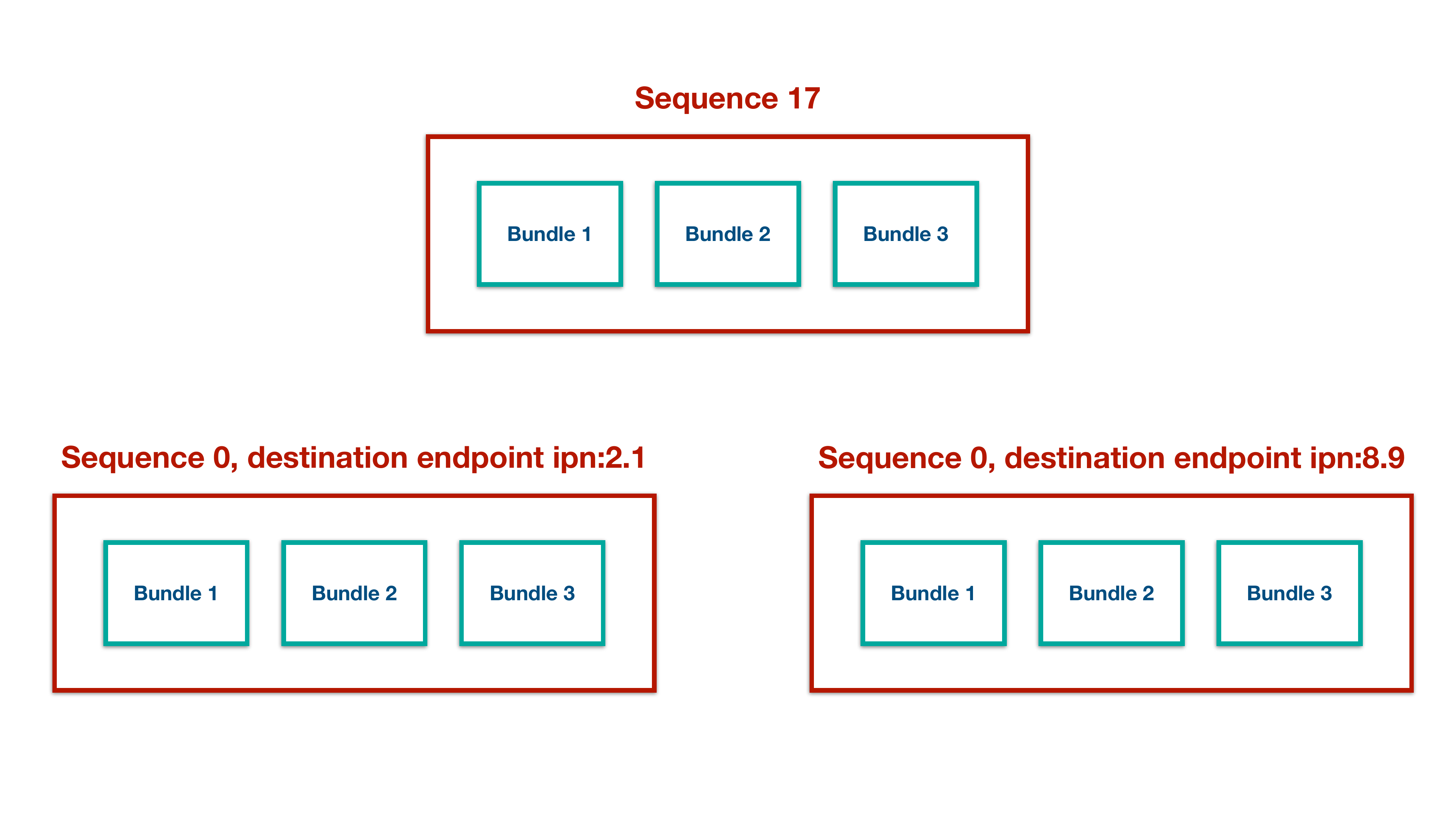}
    \caption{Example of 3 sequences maintained on one Bundle Node, uniquely identified on the node's scope.
    On the top: 3 bundles belonging to the sequence of ID 17. On the bottom left: 3 bundles belonging to the sequence of ID 0, which is therefore uniquely identified by the destination endpoint ID of its bundles \texttt{ipn:2.1}. On the bottom right: example of 3 bundles belonging to the sequence of ID 0, which is therefore uniquely identified by the destination endpoint ID of its bundles \texttt{ipn:8.9}.}
    \label{fig:two_cases}
\end{figure}

Maintaining several sequences with different IDs allows for high flexibility. For example, bundles used to transmit the ADUs generated by one specific client can be identified by a specific Bundle Sequence ID. Similarly, bundles routed toward the same next hop can be assigned to the same sequence. However, the most common case is identifying a specific traffic by its destination endpoint. The special Bundle Sequence ID ``0” is reserved for that purpose: it indicates that separate Bundle Sequence Counters are maintained per destination endpoint on the Bundle Node. Two cases must therefore be distinguished to know which parameters will uniquely identify a specific bundle in the network:
\begin{itemize}
    \item If the Bundle Sequence ID is not 0, then (1) the Bundle Sequence Number, (2) the Bundle Sequence ID, and (3) the Block Source Administrative Endpoint ID will uniquely identify the bundle (see Figure~\ref{fig:two_cases}, top).
    
    \item If the Bundle Sequence ID is 0, then (1) the Bundle Sequence Number, (2) the Bundle Destination Endpoint ID, and (3) the Block Source Administrative Endpoint ID will uniquely identify the bundle (see Figure~\ref{fig:two_cases}, bottom left and right).
\end{itemize}

This way, we ensure no two bundles in the network will have the same tuple \{[Bundle Sequence ID / Destination Endpoint ID], Bundle Sequence Number, 
Block Source Administrative Endpoint ID\}. Once added to a bundle, the extension block(s) containing those identification properties will give subsequent nodes of the network the necessary information to report on the status of the bundle in the case of Compressed Bundle Reporting and on the status of custody transfer operations involving that bundle in the case of Compressed Custody Signalling. 

\subsection{General format of compressed reports} \label{report_format}

 Compressed Bundle Reporting and Compressed Custody Signalling use two new Administrative Records to report on the status of bundles: the Compressed Reporting Signal (CRS) and the Compressed Custody Signal (CCS). Two fundamental objects were introduced to report on a set of bundles concisely: bundle sequences and bundle sequence collections. They utilize the identification elements introduced in Section~\ref{unique_id} to compress the report information as much as possible. 
\newline

A Bundle Sequence represents a range of Bundle Sequence Numbers and is defined as a definite length Concise Binary Object Representation (CBOR) array of 3 or 4 elements:

\begin{itemize}
    \item the \textbf{First Sequence Number} (CBOR unsigned int)
    \item the \textbf{Sequence Length} (CBOR unsigned int)
    \item the \textbf{Bundle Sequence ID} (CBOR unsigned int) if it is not equal to 0, else the Destination Endpoint ID
    \item optionally, the \textbf{Block Source Administrative Endpoint ID}. It can be omitted if it is equal to the Administrative Endpoint ID of the node receiving the report, like for all Compressed Custody Signals, which always omit it (see Section~\ref{ccs_section}).
\end{itemize}

Future improvements to the bundle sequence object may include adding optional time span information. Depending on the type of report, this could, for example, include the time of reception, forwarding, or delivery of the first and last bundles of the sequence. 
\newline

\begin{figure}
    \centering
    \includegraphics[width=0.7\linewidth]{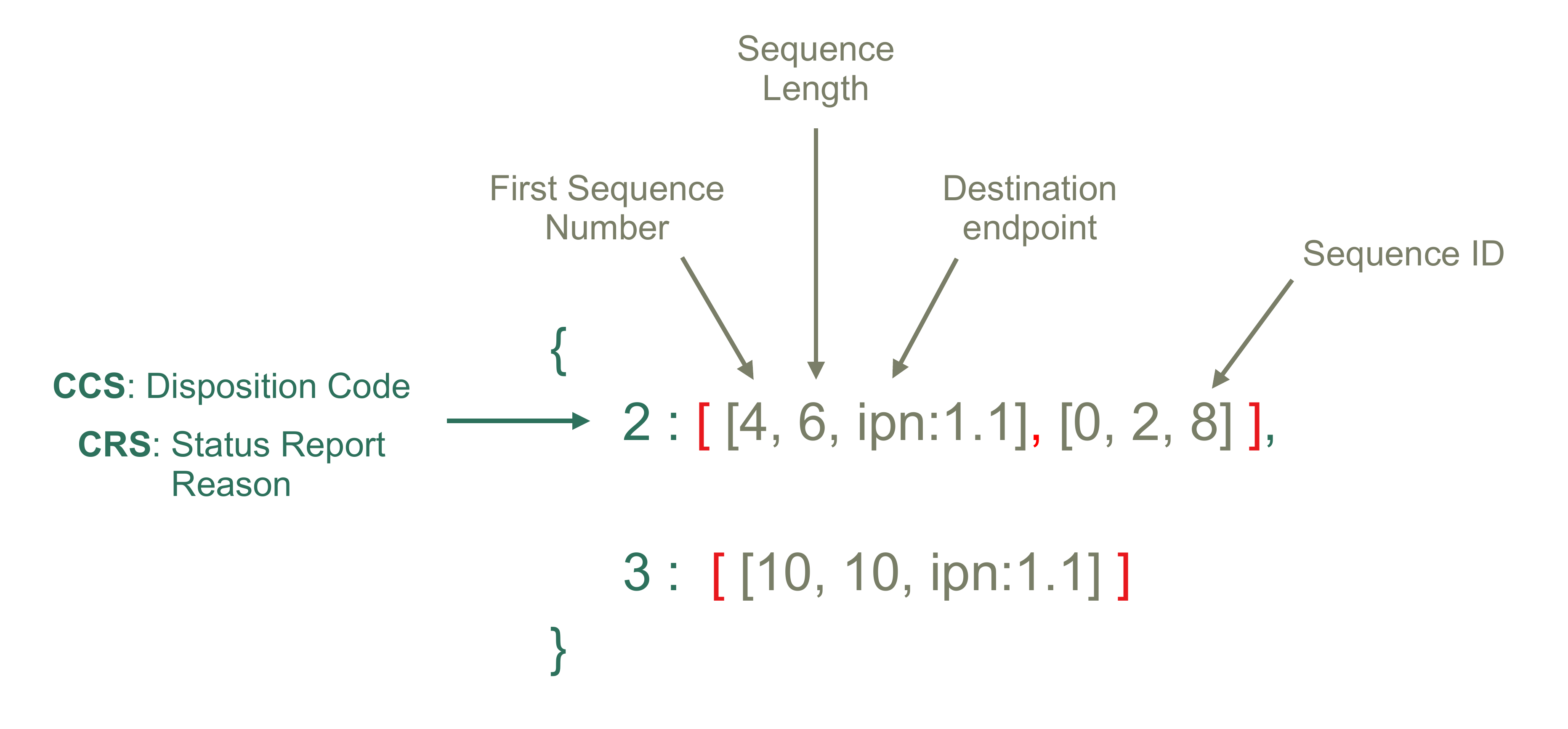}
    \caption{Example of a reporting signal for Compressed Bundle Reporting or Compressed Custody Signalling. A Bundle Sequence Collection is represented by red brackets, and a Bundle Sequence by gray brackets. On the first line, the Bundle Sequence Collection consists of two Bundle Sequences, one describing bundles 4 to 9 in the sequence related to destination endpoint ID \texttt{ipn:1.1}, and one describing bundles 0 to 1 in the sequence of ID 8. }
    \label{fig:ccs_crs}
\end{figure}

A Bundle Sequence Collection is a definite length CBOR array of Bundle Sequences. They are used to build the content of the two newly defined Administrative Records by mapping a Bundle Sequence Collection to a specific status report reason (deletion, reception, forwarding, delivery, etc.) in the case of Compressed Bundle Reporting and to a specific disposition code (custody accepted, custody refused, etc.) in the case of Compressed Custody Signalling. An example of this is given in Figure~\ref{fig:ccs_crs}, and the available status report reasons and disposition codes will be detailed respectively in Section~\ref{cbr_section} and Section~\ref{ccs_section}.

\subsection{Compressed Bundle Reporting} \label{cbr_section}

The Compressed Bundle Reporting mechanism is as follows. First, a node (not necessarily the bundle's source) will request a specific type of reporting for a bundle. Second, it will give the bundle a Bundle Sequence ID and a Bundle Sequence Number, and both information will be stored in a Compressed Reporting Extension Block added to the bundle. Third, the node will forward the bundle, and subsequent nodes in the path will generate Compressed Reporting Signals, if necessary, based on the type of reporting required. They will forward those CRSs to the report destination specified in the CREB. A CREB will not be modified while the bundle travels through the network, but subsequent nodes can add more CREBs if they need a specific type of reporting.
\newline

\begin{figure}
    \centering
    \includegraphics[width=0.7\linewidth]{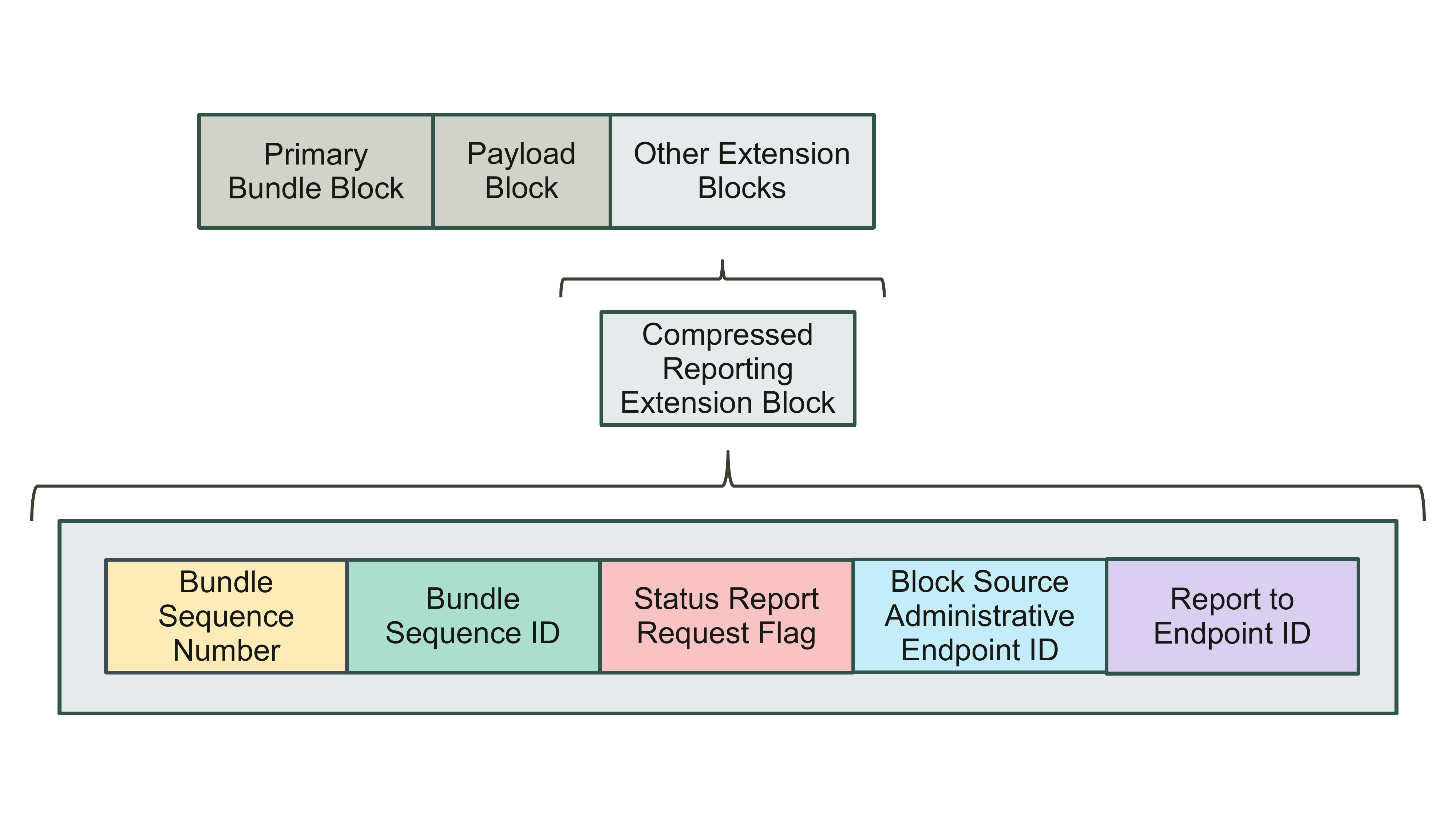}
    \caption{Format of the Compressed Reporting Extension Block.}
    \label{fig:cbr_block}
\end{figure}

A CREB (depicted in Figure~\ref{fig:cbr_block}) is an array of up to five fields. Including a field in the array requires the inclusion of all previous fields in the specified order.

\begin{itemize}
    \item \textbf{The Bundle Sequence Number}. This is the only mandatory field. A CREB containing only a Bundle Sequence Number can be used for gap detection or in-sequence delivery of bundles. 
    
    \item \textbf{The Bundle Sequence ID}. If omitted during encoding by the Block Source Node, it will default to 0 during the decoding procedure on subsequent nodes. In this case and as explained in Section~\ref{unique_id}, the bundle belongs to the sequence specific to its destination endpoint ID maintained on the Block Source Node.

    \item \textbf{The type of reporting required} (reception, deletion, forwarding, delivery). If absent, no reporting will occur.

    \item \textbf{The Block Source Administrative Endpoint ID}. It can be omitted if the node that inserted the CREB is also the bundle's source. In that case, it will default to the bundle's Source Node ID in the Primary Block during the decoding procedure on subsequent nodes.

    \item \textbf{The Report Endpoint ID}. If specified, the reports will be sent there. If not, they will be sent to the Block Source Administrative Endpoint ID if it is specified. Otherwise, reports will be destined to the Source Node ID in the Primary Block.

\end{itemize}

A node receiving a bundle with a CREB will report on the bundle's status according to the requested report type after aggregating several reports. For example, if the report type is set to deletion, then the node will only issue a CRS if it deletes the bundle at some point during bundle processing. A CRS is defined as a CBOR Map, with report types (CBOR unsigned integer) as keys and Bundle Sequence Collections as values (see Figure~\ref{fig:ccs_crs}). The available status report reasons are the following: 
\begin{itemize}
    \item 0: Reception report
    \item 1: Forwarding report
    \item 2: Delivery report
    \item 3: Deletion report
\end{itemize}

More reasons for status reports can be defined in the future.
\newline

A CRS is destined for only one node. As long as a CRS has not been sent, it is said to be pending and can be modified: new bundles whose status needs to be communicated to the destination node of the CRS can be added. The bundle node can decide to send a CRS based on two criteria: the number of bundles reported in the CRS and the time that has passed since the CRS was created. This second criterion ensures a CRS is not pending for too long, which could lead to the transmission of outdated information or incorrect assumptions from the source node awaiting the report (like a bundle presumed lost before the report is received). The maximum number of bundles and the maximum time before sending a CRS should be configurable by the Management Information Base (MIB) network operator. 

Upon receiving a CRS, additional reactive behaviours on the receiving node could be implemented: a node could, for example, perform retransmission if it receives a bundle deletion report or if delivery reporting indicates that certain bundles have not been delivered.

\subsection{Compressed Custody Signalling} \label{ccs_section}

\begin{figure}[h]
    \centering
    \includegraphics[width=0.6\linewidth]{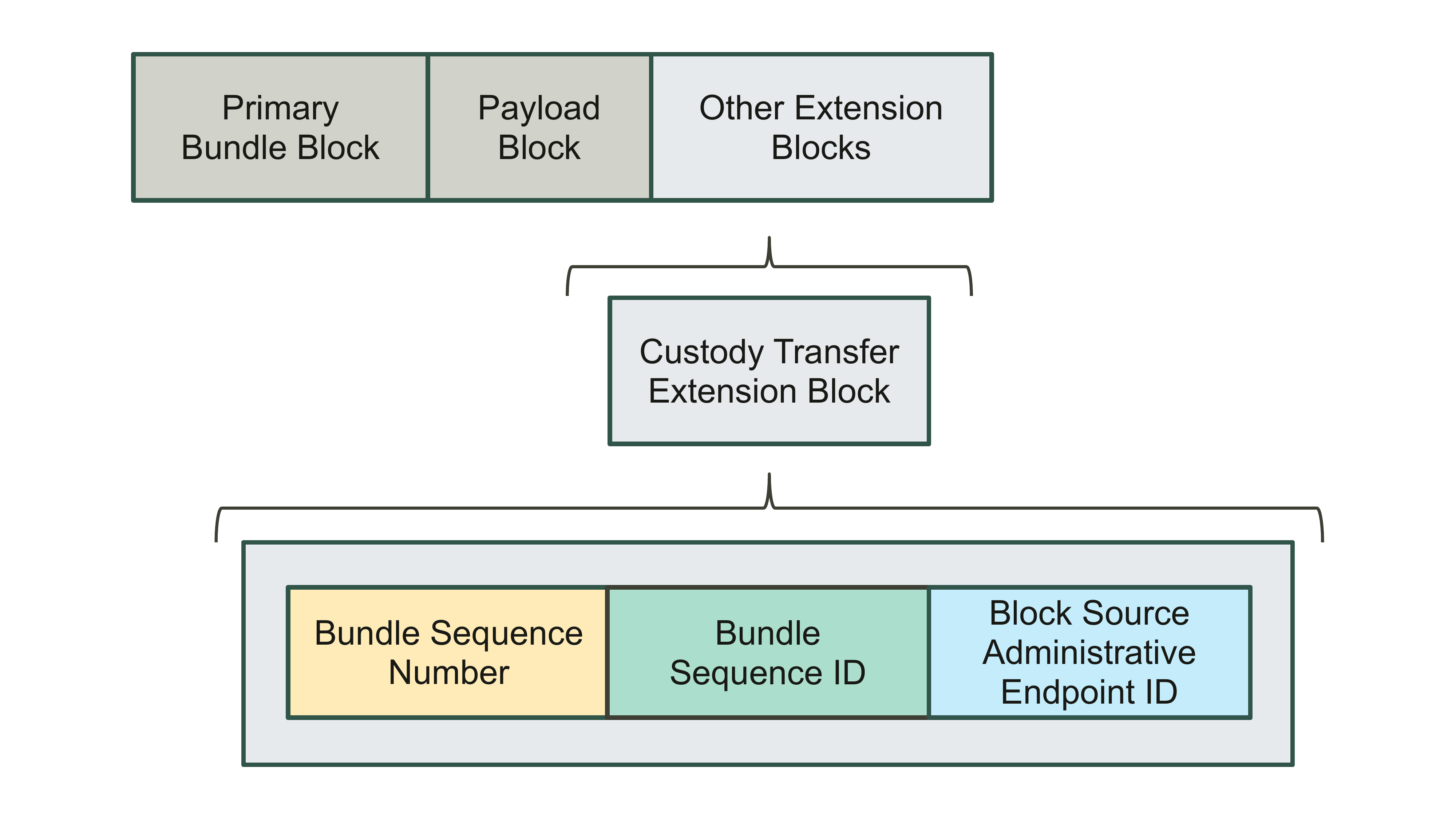}
    \caption{Format of the Custody Transfer Extension Block.}
    \label{fig:ccs_block}
\end{figure}

\begin{sidewaysfigure}
    \centering
    \includegraphics[width=1\linewidth]{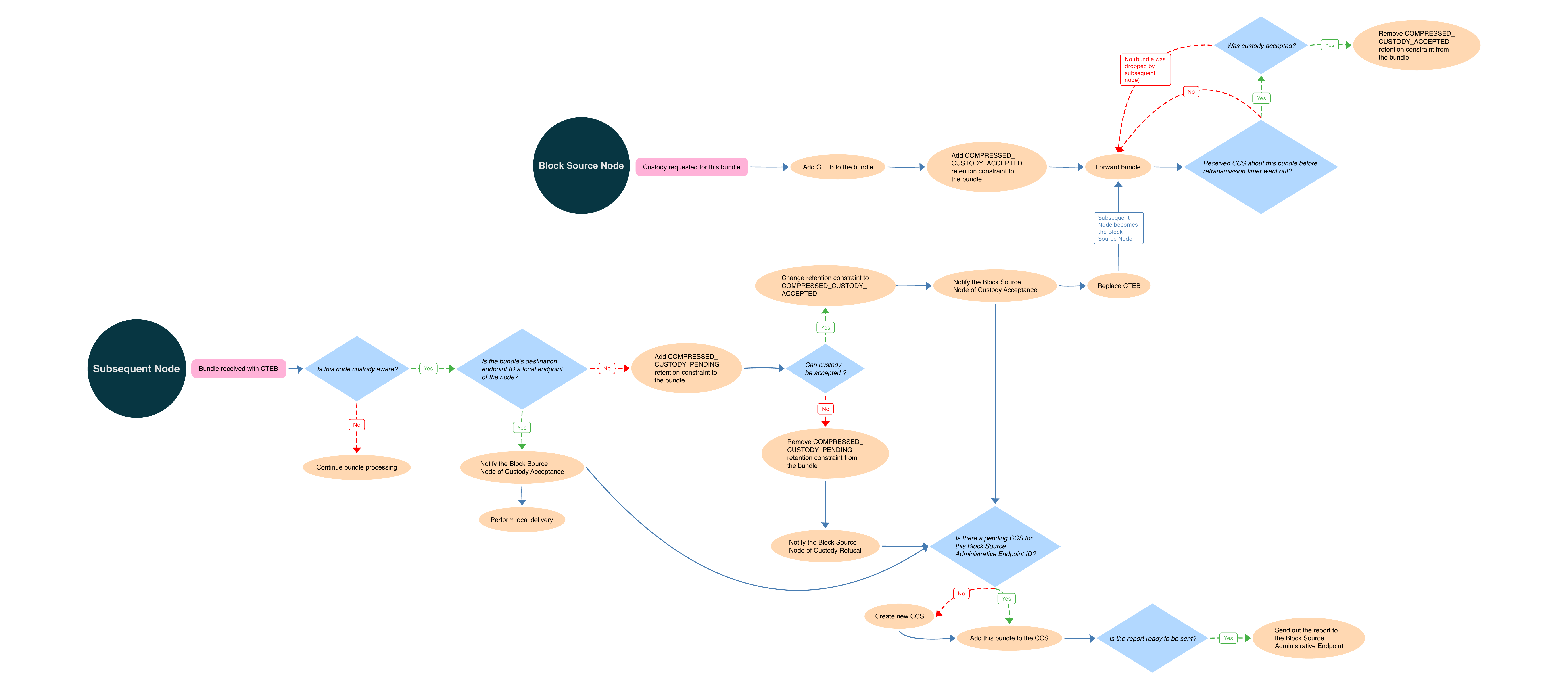}
    \caption{Flow chart of the Compressed Custody Signalling mechanism implemented in ESA BP. To keep the figure from becoming too complex, the case where custody is refused but the bundle is still forwarded was not included.}
    \label{fig:ct-flow-chart}
\end{sidewaysfigure}

The mechanism of Compressed Custody Signalling implemented in ESA's Bundle Protocol implementation is presented in Figure~\ref{fig:ct-flow-chart}. A node starts by requesting Compressed Custody Signalling for a specific bundle and inserts a Custody Transfer Extension Block, depicted in Figure~\ref{fig:ccs_block}. This block is similar to the CREB since it contains:
\begin{itemize}
    \item the \textbf{Bundle Sequence Number}
    \item the \textbf{Bundle Sequence ID}
    \item the \textbf{Block Source Administrative Endpoint ID}
\end{itemize}

It differs from the CREB for the following reasons: (1) all three fields are mandatory, (2) there can only be one CTEB per bundle, and (3) there is no report endpoint ID specified. Reports are always sent to the Block Source Administrative Endpoint ID. 
\newline

Once a bundle with a CTEB has been forwarded, the Block Source Node will wait for a Compressed Custody Signal. If the node receives no CCS reporting on the bundle, it will be retransmitted. Retransmission decisions can be triggered by several events, such as a timer going off (the bundle gets retransmitted after a certain amount of time if no related custody success signal has been received), gap detection using Custody Acceptance signals, a contact start, an explicit command to retransmit specific bundles currently under custody, a forwarding failure reported by the Convergence Layer Adapter, etc. 

To allow for retransmissions, as long as the node is the current custodian of a bundle and the bundle's lifetime has not expired, it needs to have the bundle saved in its persistent memory. The way to do this in BP is to use a retention constraint\footnote{\cite{RFC9171} defines a retention constraint as "an element of the state of a bundle that prevents the bundle from being discarded. That is, a bundle cannot be discarded while it has any retention constraints”.}. Two new ones were implemented in ESA BP: COMPRESSED CUSTODY PENDING (used while the Bundle Protocol Agent is deciding if it can accept custody for the bundle) and COMPRESSED CUSTODY ACCEPTED (used for as long as the node is the current custodian of the bundle).

Any subsequent node receiving a bundle with a CTEB will evaluate custody based on resources (storage, available route to the destination, timely contact with the next node on the path, etc.) and node-specific policies. A node can accept or refuse custody and, in the latter case, either drop the bundle or forward it to the next hop. If it decides to accept custody and becomes the new custodian, it should become the new recipient of any future Compressed Custody Signals concerning this bundle. For this to happen, the node will replace the CTEB in the bundle. Based on its policies, it will give a new identity (Bundle Sequence Number and Bundle Sequence ID) to the bundle and specify its Administrative Endpoint ID as the Block Source Administrative Endpoint ID \footnote{This last point explains why no report endpoint ID needs to be specified in the CTEB, and why the Block Source Administrative Endpoint ID is always omitted in a Compressed Custody Signal: the reports will always be destined to the current custodian (the Block Source Node) since each time a new node accepts custody, a new CTEB is created and replaces the old one.}. 
\newline

The custody decision will be added to a Compressed Custody Signal, sent using the same criteria as Compressed Reporting Signals (see Section~\ref{cbr_section}). The format of a CCS is also very similar to that of a Compressed Reporting Signal (see Figure~\ref{fig:ccs_crs}). A CCS is defined as a CBOR Map, with disposition codes (CBOR unsigned integer) as keys and Bundle Sequence Collections as values. The available disposition codes are the following:
\begin{itemize}
\item For Custody Acceptance, a positive CBOR integer:
    \begin{itemize}
    \item [*] 1: Custody Accepted (no further information). The current custodian should delete the bundle from its persistent memory by removing the retention constraint.
    
    \item [*] 2: Custody Accepted (duplicate reception, custody was already accepted for this bundle). The current custodian should delete the bundle from its persistent memory by removing the retention constraint, and adapt its retransmission timers to prevent further redundant reception for future bundles.

    \item [*] Values above 2: reserved for future use.
    \end{itemize}
\item For Custody Refusal, a negative CBOR integer:
    \begin{itemize}
    \item [*] -1: Custody Refused (bundle dropped). The current custodian should retransmit the bundle and stay custodian.
    
    \item [*] -2: Custody Refused (bundle forwarded). The current custodian should reset the bundle's retransmission timer and stay custodian.

    \item [*] Values below -2: reserved for future use.
    \end{itemize}
\end{itemize}

\section{Experiments and results}

\begin{figure}[h]
    \centering
    \includegraphics[width=0.8\linewidth]{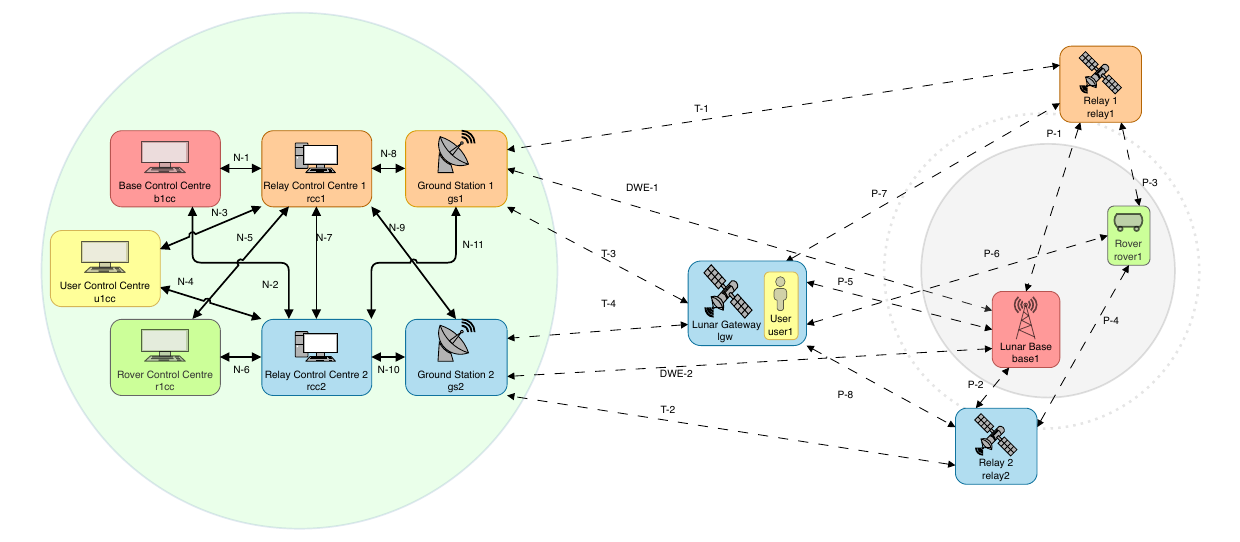}
    \caption{Lunar Communication reference scenario for Delay Tolerant Networks \cite{referenceScenarios}.}
    \label{fig:lc-config}
\end{figure}

\begin{figure}[h]
    \centering
    \includegraphics[width=0.6\linewidth]{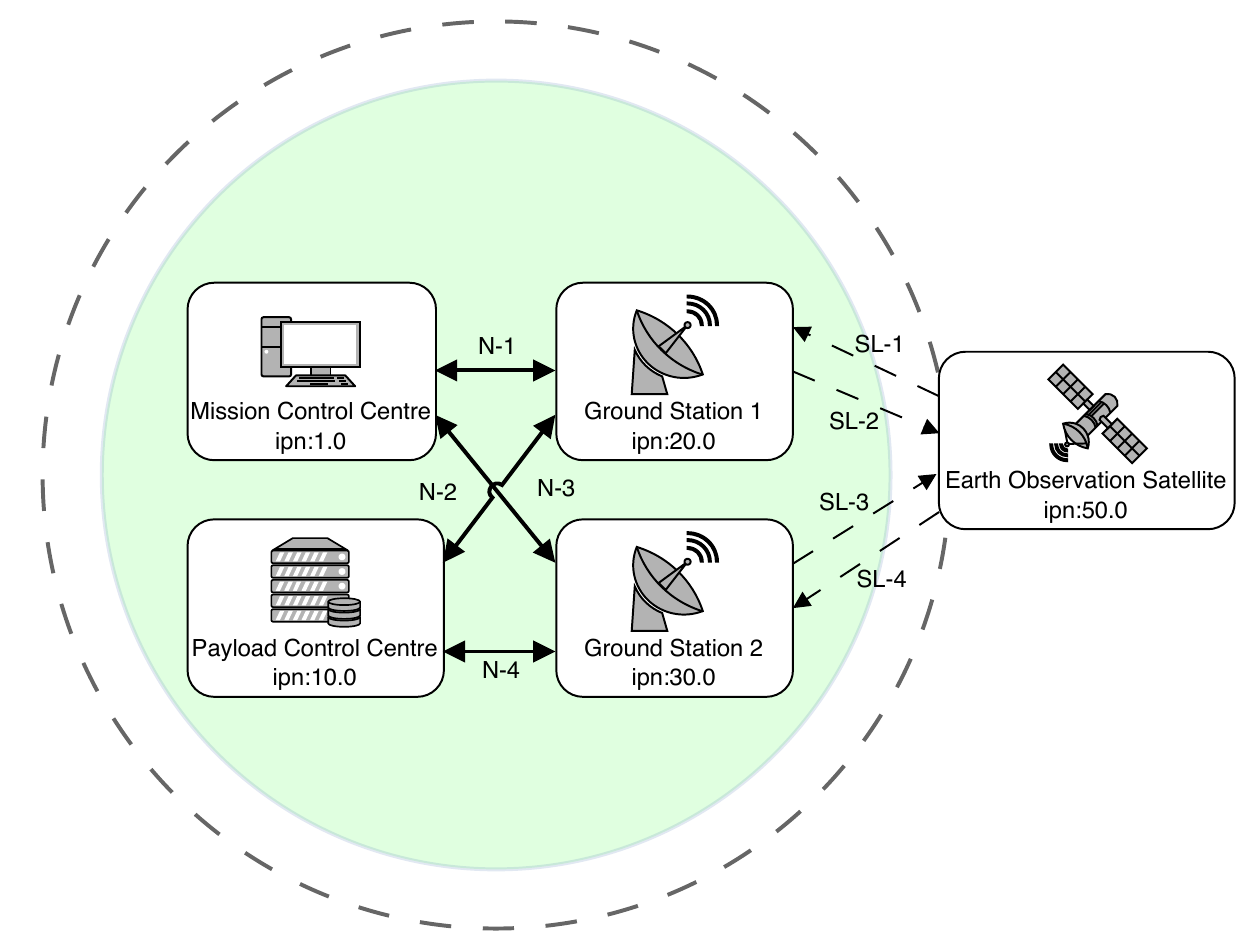}
    \caption{Earth Observation reference scenario for Delay Tolerant Networks \cite{referenceScenarios}.}
    \label{fig:eo-config}
\end{figure}

This section describes the experiments conducted with the ESA BP implementation to demonstrate the functioning of the two extensions. The experiments focused on two services: efficient end-to-end (E2E) accounting with Compressed Bundle Reporting and custody transfer with Compressed Custody Signalling. It was carried out using two network configurations developed by the authors of \cite{referenceScenarios}: a realistic Lunar Communication reference scenario (see Figure \ref{fig:lc-config}) and a realistic Earth Observation scenario (see Figure \ref{fig:eo-config}). Those scenarios were run with the Network Simulation \& Emulation Environment (NSE$^2$), a framework developed by ESA.


\subsection{Demonstration of Compressed Bundle Reporting in a realistic Lunar Communication scenario}
\label{sec:cbr-demo}

The demonstration of Compressed Bundle Reporting was performed in the Lunar communication scenario. Fifty bundle requests destined to client \texttt{ipn:21.1} (connected to \textit{rover1}) were generated by client \texttt{ipn:31.1} (connected to \textit{user1} on the Lunar Gateway) with Compressed Bundle Reporting activated for delivery events. They all transited through the Lunar Gateway (\textit{lgw}, \texttt{ipn:220.0}) and were then sent directly to \textit{rover1} since the two nodes had direct contact for the entire duration of the experiment. A packet loss of 1\% was configured on the link between \textit{lgw} and \textit{rover1}. All nodes had a limit of 100 bundles per Compressed Reporting Signal and were configured to send any pending Compressed Reporting Signal after 10 seconds if the bundle threshold was not reached.

\begin{figure}[h]
    \centering
    \includegraphics[width=1\linewidth]{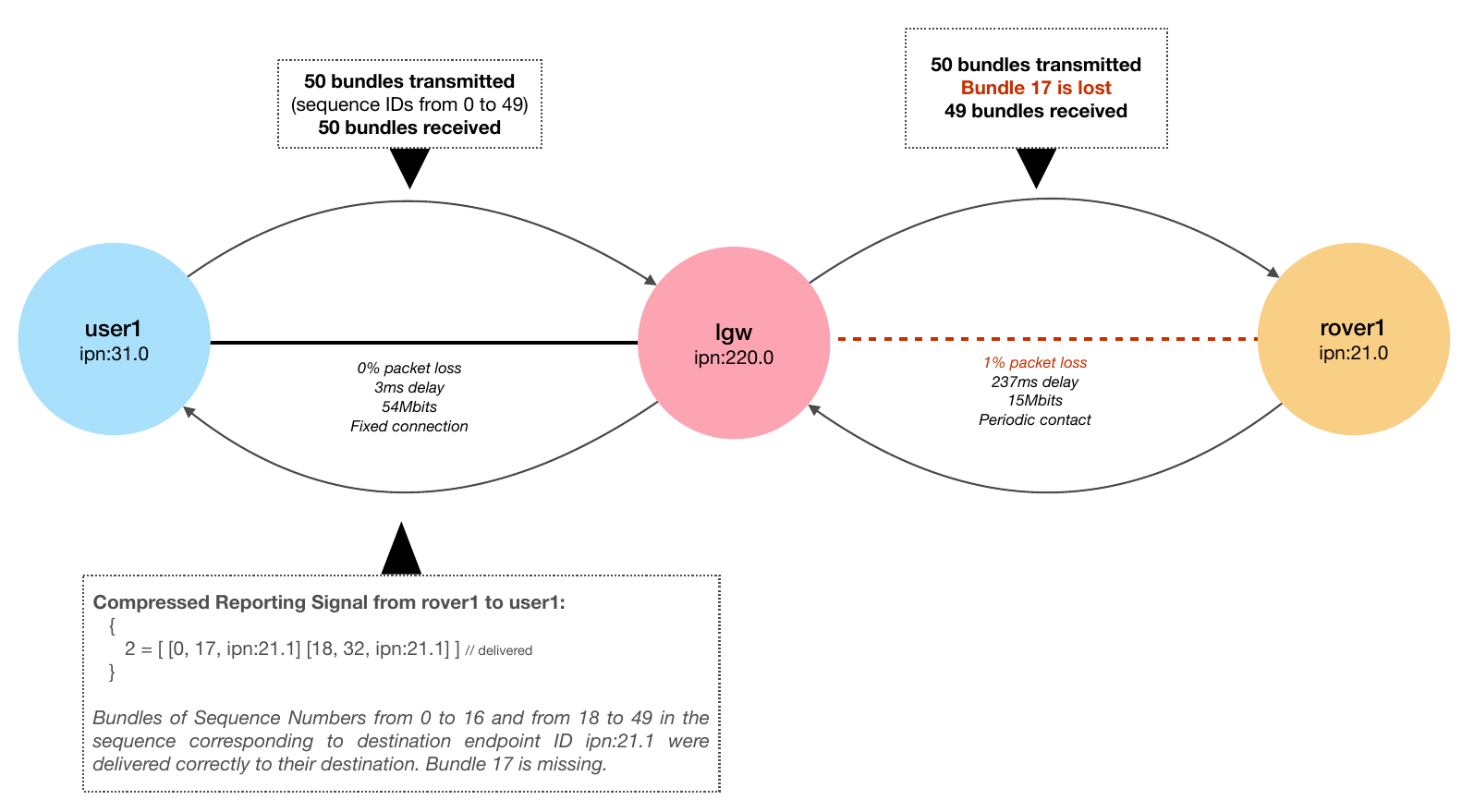}
    \caption{Overview of the events unfolding in the network during the demonstration of Compressed Bundle Reporting in a realistic Lunar Communication scenario. All nodes were configured to maintain one Bundle Sequence Counter per destination endpoint ID. This means that a bundle is identified using its Sequence Number inside the sequence maintained by its source node (here \textit{user1}) for its destination endpoint ID (here \texttt{ipn:21.1}, connected to \textit{rover1}).}
    \label{fig:cbr-lc}
\end{figure}

Figure \ref{fig:cbr-lc} shows the events that unfolded in the relevant part of the network upon transmission of the 50 bundles. In total, 49 were delivered to \textit{rover1}, and 1 (bundle of Sequence Number 17) was lost on the link between \textit{lgw} and \textit{rover1}. This led to a Compressed Reporting Signal generated by \textit{rover1} and destined to \textit{user1} containing a gap in its sequence. The transmission of the CRS was triggered when its maximum pending time was reached since the amount of bundles was below the threshold.

With this CRS, \textit{user1} can assume a delivery failure for Bundle 17. The node could be configured to retransmit the bundle to ensure E2E reliability, possibly after a timeout to make sure the missing bundle wasn't simply delivered later or was not added to a different CRS not received yet. For a more efficient reliability mechanism, Custody Transfer with Compressed Custody Signals can be used to limit retransmissions to the current custodian of a bundle, therefore avoiding E2E retransmissions.

\subsection{Demonstration of Compressed Custody Signalling in a realistic Earth Observation scenario}
\label{sec:ccs-demo}

\begin{figure}
    \centering
    \begin{subfigure}{1\textwidth}
        \centering
        \includegraphics[width=\linewidth]{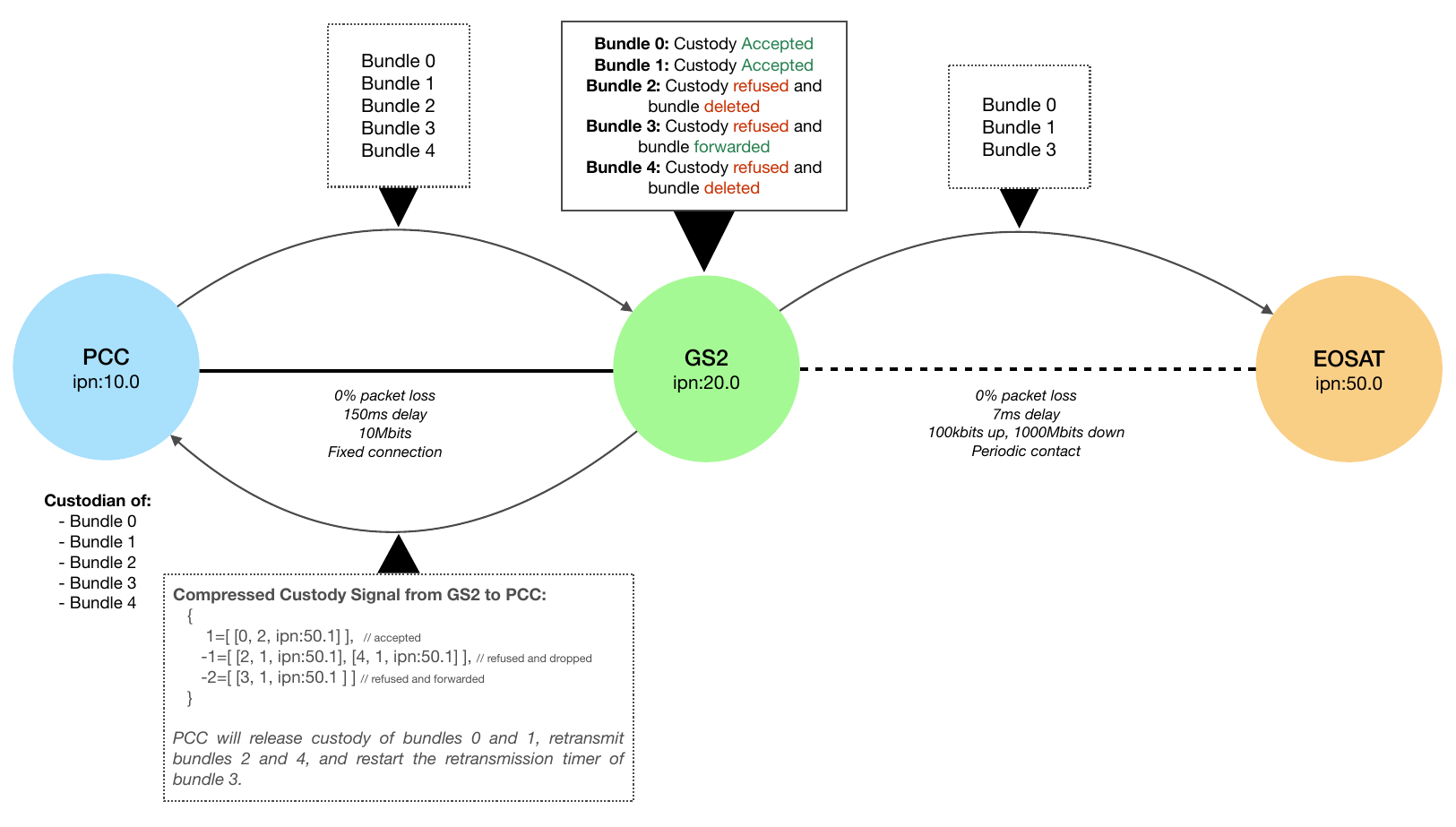}
        \caption{First phase.}
    \end{subfigure}
    \par\bigskip
    \begin{subfigure}{1\textwidth}
        \centering
        \includegraphics[width=\linewidth]{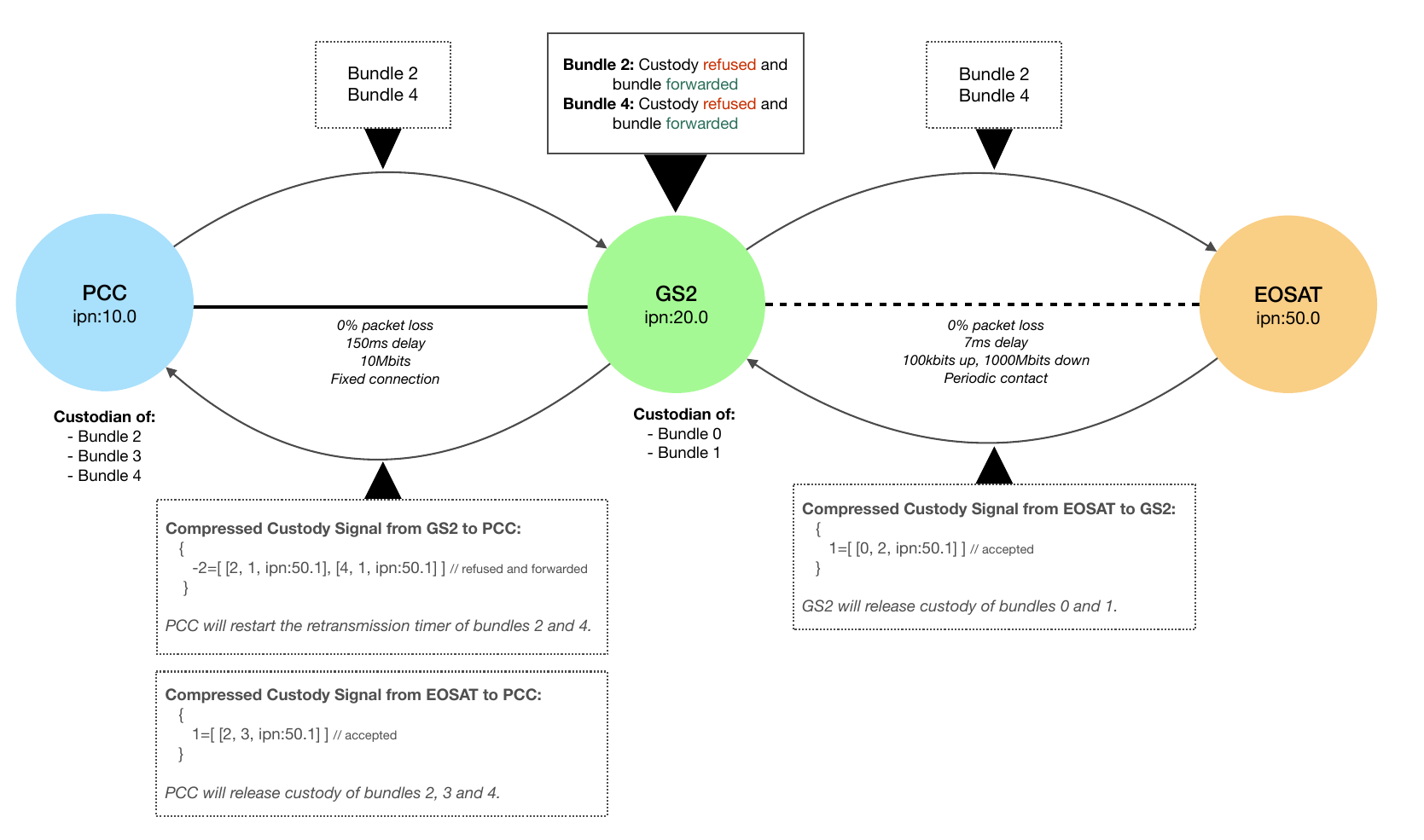}
        \caption{Second phase.}
    \end{subfigure}
    \caption{Overview of the events unfolding in the network during the demonstration of Compressed Custody Signalling in a realistic Earth Observation scenario. All nodes were configured to maintain one Bundle Sequence Counter per destination endpoint ID. This means a bundle is identified using its Sequence Number inside the sequence maintained by its current custodian for its destination endpoint ID (here, ipn:50.1).}
    \label{fig:ccs-events-eo}
\end{figure}

The demonstration of Compressed Custody Signalling was performed in the Earth Observation scenario. Five bundles were transmitted from the Payload Control Center (PCC, \texttt{ipn:10.0}) to the Earth Observation Satellite (EOSAT, \texttt{ipn:50.0}) with Compressed Custody Signalling activated. The bundle requests were generated by client \texttt{ipn:10.1} on PCC and destined to client \texttt{ipn:50.1} on EOSAT. They all transited through Ground Station 2 (GS2, \texttt{ipn:20.0}), the only ground station to have active contact with the satellite during the whole duration of the experiment. 

The path followed by the packets included only one node between the source and the destination, in this case, GS2, meaning that it was the only node that needed to take a custody decision. Indeed, PCC and EOSAT were the bundles' source and destination, and custody was always accepted. For testing purposes, GS2 was programmed to accept custody 50\% of the time and to refuse it otherwise with two possible policies: dropping the bundle (25\% chance) or forwarding it without taking custody (25\% chance). All nodes had a maximum of five bundles per Compressed Custody Signal, meaning a CCS was immediately sent when five bundles were added. They were also configured to send any pending Compressed Custody Signal after 15 seconds if the bundle threshold was not reached.

The events that unfolded in the relevant part of the network after all five bundles were transmitted are summarized in Figure \ref{fig:ccs-events-eo}. The two phases were as follows:

\begin{itemize}
    \item In the first phase, GS2 accepted custody for bundles \{0, 1\}, and refused it for bundles \{2, 3, 4\}, dropping bundles \{2, 4\} and forwarding bundle 3 to EOSAT. A Compressed Custody Signal was sent immediately from GS2 to PCC since the threshold of 5 bundles was reached.
    \item In the second phase, PCC retransmitted the deleted bundles \{2, 4\}. GS2 refused custody once again, but the two bundles were forwarded to EOSAT this time. At that point, three CCSs were pending on the different nodes: 
    one on GS2 (destined to PCC for bundles \{2, 4\}) and two on EOSAT (one destined to PCC for bundles \{2, 3, 4\}, and one destined to GS2 for bundles \{0, 1\}). As they never reached their bundle threshold, those CCS were transmitted when their maximum pending time of 15 seconds was reached. 
\end{itemize}

All bundles were correctly delivered to EOSAT less than 2 seconds after the beginning of the experiment. PCC and GS2 released custody for the bundles they were responsible for about 17 seconds after the beginning of the experiment upon reception of the last Compressed Custody Signals, which successfully terminated the exchange. Those 17 seconds correspond to the 2 seconds taken to deliver all bundles, added to the time limit of 15 seconds, after which pending Compressed Custody Signals that hadn't reached their bundle threshold were to be sent. The exchange could have been terminated faster by lowering the time limit. Generally, the maximum pending time and the bundle threshold should be adapted to the network configuration for better performance. Moreover, the amount of bundles sent in this experiment was purposefully low to keep the number of events to a level that could be easily visualized in a diagram. The relevance of using Compressed Custody Signals spanning a list of bundles instead of individual custody signals becomes clearer as the number of bundles grows. 

\subsection{Summary of Experimental Results}

The experimental results for Compressed Bundle Reporting and Compressed Custody Signalling were obtained using two realistic reference scenarios: a Lunar Communication scenario and an Earth Observation scenario, respectively. Table \ref{tab:results-summary} summarizes the input parameters used and the key performance metrics observed during the experiments.

\begin{table}[h]
\centering
\begin{tabular}{|p{5cm}|>{\centering\arraybackslash}p{5cm}|>{\centering\arraybackslash}p{5cm}|}
\hline
\centering \textbf{Metrics/Parameters}                  & \textbf{Lunar Communication Scenario \newline (Compressed Bundle Reporting)} & \textbf{Earth Observation Scenario \newline (Compressed Custody Signalling)} \\ \hline
Number of Bundles Sent           & 50                                          & 5                                         \\ \hline
Bundles Delivered                & 49                                          & 5                                         \\ \hline
Bundles Lost/Dropped              & 1 (lost on a link)                      & 2 (dropped by GS2)                        \\ \hline
CRS/CCS Generation Trigger        & 10 seconds or 100 bundles                   & 15 seconds or 5 bundles                   \\ \hline
Custody Accepted (GS2)           & N/A                                         & 50\% of cases                             \\ \hline
Custody Refused (GS2)                  & N/A                                         & 50\% of cases                             \\ \hline
Last Bundle Delivery Time               & N/A                       & \textless{} 2 seconds                     \\ \hline
Final Custody Release Time              & N/A                                         & 17 seconds after experiment start         \\ \hline
\end{tabular}
\caption{Summary of Experimental Results for Compressed Bundle Reporting and Compressed Custody Signalling.}
\label{tab:results-summary} 
\end{table}

In the Lunar Communication scenario, 50 bundles were transmitted from \textit{user1} to \textit{rover1}, with Compressed Bundle Reporting enabled for delivery events. Due to a 1\% packet loss configured on the link, 49 bundles were successfully delivered, while one bundle was lost. A Compressed Reporting Signal containing a gap in its bundle sequence was generated by \textit{rover1} and sent to \textit{user1} after 10 seconds, as the bundle threshold of 100 was not reached.

In the Earth Observation scenario, five bundles were transmitted from the Payload Control Center to the Earth Observation Satellite via Ground Station 2. Ground Station 2 was configured to accept custody in 50\% of cases, while the remaining 50\% resulted in either dropping or forwarding the bundle without custody. A Compressed Custody Signal was sent after five bundles were accumulated or 15 seconds elapsed. After several bundles were dropped and retransmitted, all bundles were correctly delivered to EOSAT within 2 seconds, and custody was released for all bundles in 17 seconds.

\section{Conclusions}

This paper defined two extensions for BPv7: Compressed Bundle Reporting and Compressed Custody Signalling. 

The first extension provides an acknowledgment mechanism using status reports which contain information about the state of a collection of bundles currently progressing in the network. Nodes can request to get notified regarding specific bundle events like delivery, deletion, forwarding or reception. Compressed Bundle Reporting offers the same functionalities as Bundle Status Reporting, but allows to report on a set of bundles using only one Administrative Record, while Bundle Status Reporting generates one Administrative Record per bundle. This effectively reduces
the amount of reports flowing in the network and helps avoid the infamous flooding often observed when using Bundle Status Reporting. 

The second extension provides an efficient reliability mechanism for BPv7. It is based on Custody Transfer, which moves the retransmission responsibility of a bundle along its path. This prevents a bundle lost at the very end of its route from getting retransmitted by the source, end-to-end retransmissions being a very inefficient behaviour in the context of space networks with intermittent contacts and high latencies. Compressed Custody Signalling implements the
Custody Transfer mechanism and makes use of the same Administrative Record format as Compressed Bundle Reporting to avoid flooding the network with reports. One report summarises the states of Custody Transfer operations for a set of bundles, an
improvement compared to the Custody Transfer mechanism defined for BPv6 which generates one Administrative Record per bundle.

Each mechanism uses a new extension block (Compressed Reporting Extension Block and Custody Transfer Extension Block) which is added to a bundle to provide it with a unique identifier within the scope of the network, allowing subsequent nodes to report on its state. Each mechanism also provides a new Administrative Record (Compressed Reporting Signal and Compressed Custody Signal).

The two extensions were tested and validated in reference Delay Tolerant Network scenarios, including a Lunar Communication and an Earth Observation scenario. The tests focused on two crucial services for networks: end-to-end accounting with Compressed Bundle Reporting, and efficient retransmissions using Compressed Custody Signalling.

The described extensions are expected to be standardised in a CCSDS Experimental Specification with some minor modifications in 2025. Further work will validate Custody Transfer and Compressed Bundle Reporting in more complex scenarios and analyse different options for bundle retransmission. Another focus will be on the realisation of E2E services such as duplicate suppression, in-sequence delivery, E2E accounting and reliability, and the use of compressed bundle status reporting for DTN Network Management purposes.

\printbibliography 

\end{document}